\documentclass[traditabstract]{aa} % for the abstract without structuration 
\usepackage{graphicx}
\usepackage{txfonts}
\def  \cmsq     {\ifmmode {\rm cm}^{-2} \else cm$^{-2}$\fi}
\def  \ergs     {\ifmmode {\rm erg\,s}^{-1} \else erg s$^{-1}$\fi}
\def  \ergcms   {\ifmmode {\rm erg\,cm}^{-2}\,{\rm s}^{-1}
                        \else erg\,cm$^{-2}$\,s$^{-1}$\fi}
\def \lhard  {\ifmmode {\rm L_{2-10keV}} \else ${\rm L_{2-10keV}}$\fi}
\def \lir  {\ifmmode {\rm L_{IR}} \else ${\rm L_{IR}}$\fi}
\def \nh  {\ifmmode {\rm N_{H}} \else ${\rm N_{H}}$\fi}
\def \Msun {\ifmmode M_{\odot} \else $M_{\odot}$\fi}
\def \Lsun {\ifmmode L_{\odot} \else $L_{\odot}$\fi}
\def \spitzer  {{\it Spitzer}}

\def \herschel {{\it Herschel}}

\def \iso      {{\it ISO}}

\begin{document}
   \title{PACS Evolutionary Probe (PEP) - A Herschel Key Program
  \thanks{Herschel is an 
  ESA space observatory with science instruments provided by European-led 
  Principal Investigator consortia and with important participation from 
  NASA.} }

   \subtitle{}

   \authorrunning{Lutz et al.}
   \titlerunning{PACS evolutionary probe}

   \author{D.~Lutz\inst{1}
           \and A.~Poglitsch\inst{1}
           \and B.~Altieri\inst{2}
           \and P.~Andreani\inst{3,4}
           \and H.~Aussel\inst{5}
           \and S.~Berta\inst{1}
           \and A.~Bongiovanni\inst{6,7}
           \and D.~Brisbin\inst{8}
           \and A.~Cava\inst{6,7}
           \and J.~Cepa\inst{6,7}
           \and A.~Cimatti\inst{9}
           \and E.~Daddi\inst{5}
           \and H.~Dominguez-Sanchez\inst{9}
           \and D.~Elbaz\inst{5}
           \and N.M.~F\"orster~Schreiber\inst{1}
           \and R.~Genzel\inst{1}
           \and A.~Grazian\inst{10}
           \and C.~Gruppioni\inst{9}
           \and M.~Harwit\inst{8}
           \and E.~Le~Floc'h\inst{5}
           \and G.~Magdis\inst{5}
           \and B.~Magnelli\inst{1}
           \and R.~Maiolino\inst{10}
           \and R.~Nordon\inst{1}
           \and A.M.~P\'erez~Garc\'ia\inst{6,7}
           \and P.~Popesso\inst{1}
           \and F.~Pozzi\inst{9}
           \and L.~Riguccini\inst{5}
           \and G.~Rodighiero\inst{11}
           \and A. Saintonge\inst{1}
           \and M.~Sanchez~Portal\inst{2}
           \and P.~Santini\inst{10,1}
           \and L.~Shao\inst{1}
           \and E.~Sturm\inst{1}
           \and L.J.~Tacconi\inst{1}
           \and I.~Valtchanov\inst{2}
           \and M.~Wetzstein\inst{1}
           \and E.~Wieprecht\inst{1}
           }

\institute{MPE, Postfach 1312, 85741 Garching, Germany, 
\email{lutz@mpe.mpg.de}
\and European Space Astronomy Centre, Villafranca del Castillo, Spain
\and European Southern Observatory, Karl-Schwarzschild-Stra\ss e 2, 85748 Garching, Germany
\and INAF - Osservatorio Astronomico di Trieste, via Tiepolo 11, 34143 Trieste, Italy
\and IRFU/Service d'Astrophysique, B\^at.709, CEA-Saclay, 91191 
           Gif-sur-Yvette Cedex, France
\and Instituto de Astrof\'isica de Canarias, 38205 La Laguna, Spain
\and Departamento de Astrof\'isica, Universidad de La Laguna, Spain
\and Space Science Building, Cornell University, Ithaca, NY 14853-6801, USA
\and Istituto Nazionale di Astronomia, Osservatorio Astronomico di 
           Bologna, Via Ranzani 1, I-40127 Bologna, Italy
\and INAF - Osservatorio Astronomico di Roma, via di Frascati 33, 
00040 Monte Porzio Catone, Italy               
\and Dipartimento di Astronomia, Universit\'a di Padova, 35122 Padova, Italy
}           

   \date{received 19 April 2011 ; accepted 9 June 2011}

\abstract
{Deep far-infrared photometric surveys studying galaxy evolution and the
nature of the cosmic infrared background are a key strength of the \herschel\ 
mission. We describe the scientific motivation for the PACS Evolutionary Probe
(PEP) guaranteed time key program and its role in the complement of Herschel 
surveys, and the field selection which includes popular multiwavelength 
fields such as GOODS, COSMOS, Lockman Hole, ECDFS, EGS. We provide an account
of the observing strategies and data reduction
methods used. An overview of first science results illustrates the potential
of PEP in providing calorimetric star formation rates for high redshift 
galaxy populations, thus testing and superseeding previous extrapolations from
other wavelengths, and enabling a wide range of galaxy evolution studies.
}
 
%  \abstract


   \maketitle
%
%________________________________________________________________

\section{Motivation}
Over the last two decades, it has become increasingly clear that no 
understanding of galaxy evolution can be obtained without accounting for
the energy that is absorbed by dust and re-emitted at mid- and far-infrared 
wavelengths. For example, early attempts to reconstruct the cosmic star 
formation history suffered from uncertainties in the obscuration 
corrections that have to be applied to the rest frame ultraviolet
measurements (e.g. Madau \cite{madau96}, Lilly et al. \cite{lilly96}). 
Soon, the importance of 
luminous dusty high redshift galaxies was highlighted by
the detection of infrared-luminous populations both in the mid-infrared (e.g.
Aussel et al. \cite{aussel99}, Genzel \& Cesarsky \cite{genzel00}) and the 
submm (e.g. Hughes et al. \cite{hughes98}). Mainly due to the large 
mid-infrared legacy that the \spitzer\ mission provided for extragalactic
studies (Soifer et al. \cite{soifer08}), a global 
consistency between these two perspectives from the rest frame ultraviolet 
and from the rest frame mid-IR side 
could be achieved (e.g. Hopkins and Beacom \cite{hopkins06}). This is
because the rest frame ultraviolet obscuration on average can be constrained 
by comparing the observed ultraviolet emission to the sum of ultraviolet 
and infrared emission. 
Extrapolation from the mid-infrared to the rest far-infrared 
was still necessary, however, on the basis of SED assumptions that
were untested at the redshifts to which they had to be applied.

At the same time, the detection of the cosmic far-infrared background (CIB)
with total energy content similar to the optical/near-infrared one 
(Puget et al.
\cite{puget96}, Hauser et al. \cite{hauser98}) highlighted the importance 
of dust emission in the cosmic energy budget. An increase with redshift 
in the energy output of dusty galaxies relative to others was inferred 
both from the shape of the CIB and from the more rapid 
increase with redshift of IR energy density compared to ultraviolet energy 
density in the resolved observations (e.g., Le Floc'h \cite{lefloch05}). 
All these lines of evidence
strongly suggest that our picture of high redshift galaxy evolution
is substantially incomplete and emphasize the need for direct rest frame 
`calorimetric' 
far-infrared measurements of individual high-z galaxies, in order to avoid SED 
extrapolation and to increasingly replace population averages with 
individual measurements.
While small cryogenic space telescopes like \iso\ and \spitzer\ were already
equipped with sensitive far-infrared detectors, they were for these 
wavelengths rapidly limited
by source confusion, and thus focussed on the study of local objects, 
or at z$\gtrsim$0.5 on study of only the most luminous galaxies. 
They also were 
able to resolve only a small fraction of the cosmic infrared background. 

\begin{table*}
\caption{PEP fields}           
\centering                       
\begin{tabular}{lrrcrccc}      
\hline\hline              
Field        &   RA&   DEC & Size &PA&wavelengths&Observations&Time\\
     &\multicolumn{2}{c}{degree,\ J2000}&arcmin&degree&$\mu$m&&hours\\  
\hline                       
COSMOS       &150.11917&  2.20583&85$\times$85&  0\tablefootmark{a}&100,160\tablefoottext{b}&Nov 2009 -- Jun 2010&196.9\\
Lockman Hole XMM&163.17917&57.48000&24$\times$24&  0               &100,160\tablefoottext{b}&Oct 2009 -- Nov 2009&32.1\\
EGS          &214.82229& 52.82617&67$\times$10& 40.5               &100,160\tablefoottext{b}&May 2011 -- June 2011&34.8\\
ECDFS        & 53.10417&-27.81389&30$\times$30&  0                 &100,160\tablefoottext{b}&Feb 2010 -- Feb 2011&32.8\\
GOODS-S      & 53.12654&-27.80467&17$\times$11&-11.3            &70,100,160\tablefoottext{b}&Jan 2010            &239.7\\
GOODS-N      &189.22862& 62.23867&17$\times$11& 41                 &100,160\tablefoottext{b}&Oct 2009            &25.8\\
Cl0024+16    &  6.62500& 17.16250& 6$\times$6 & 45\tablefoottext{c}&100,160\tablefoottext{b}&Jun 2010            & 6.2\\
Abell 370    & 39.97083& -1.57861& 4$\times$4 & 45\tablefoottext{c}&100,160\tablefoottext{b}&                    & 5.3\\
MS0451.6-0305& 73.55000& -3.01667& 4$\times$4 & 45\tablefoottext{c}&100,160\tablefoottext{b}&May 2010            & 5.3\\
Abell 1689   &197.87625& -1.34000& 4$\times$4 & 45\tablefoottext{c}&100,160\tablefoottext{b}&Jan 2011 --         &13.0\\
RXJ1347.5-1145&206.87708&-11.75250& 4$\times$4 & 45\tablefoottext{c}&100,160\tablefoottext{b}&Feb 2011           & 5.3\\
MS1358.4+6245&209.97625& 62.51000& 4$\times$4 & 45\tablefoottext{c}&100,160\tablefoottext{b}&May 2010            & 5.3\\
Abell 1835   &210.25833&  2.87889& 4$\times$4 & 45\tablefoottext{c}&100,160\tablefoottext{b}&Feb 2011            & 5.3\\
Abell 2218   &248.97083& 66.20611& 4$\times$4 & 45\tablefoottext{c}&100,160\tablefoottext{b}&Oct 2009            &10.2\\
Abell 2219   &250.08333& 46.71194& 4$\times$4 & 45\tablefoottext{c}&100,160\tablefoottext{b}&May 2010            & 5.3\\
Abell 2390   &328.40417& 17.69556& 4$\times$4 & 45\tablefoottext{c}&100,160\tablefoottext{b}&May 2010            & 5.3\\
RXJ0152.7-1357&38.17083&-13.96250&10$\times$5 & 45                 &100,160,250,350,500     &Mar 2010 -- Jan 2011& 5.8\\
MS1054.4-0321&164.25092& -3.62428&10$\times$5 &  0                 &100,160,250,350,500     &Jun 2010 -- Dec 2010& 5.8\\
\hline                               
\end{tabular}
\tablefoot{List of the six blank deep fields, ten lensing clusters and
two z$\sim$1 clusters to be observed by PEP. As of early June 2011, 
surveys for all fields
except Abell 370 and Abell 1689 are completed (see column `Observations').
Field sizes are quoted in terms of the
nominal scan length. Given the 3.5$\arcmin\times$1.75$\arcmin$ size of the PACS
arrays, coverage is unequal at the field edges starting to decrease already
inside the quoted fields, but also extending beyond them. Observing times are
as computed in HSPOT 5.2, true execution times can differ due to overhead
changes during the \herschel\ mission.\\
\tablefoottext{a} Approximate. See Sect.~\ref{sect:obsstrat} for detailed
implementation.\\
\tablefoottext{b} 250, 350, 500~$\mu$m obtained in coordinated observations 
by the HerMES key program (Oliver et al. \cite{oliver11}).\\
\tablefoottext{c} In array coordinates, true position angle on sky will 
depend on execution date of the observation.
}
\label{tab:fields}     
\end{table*}

With ESA's \herschel\ space observatory (Pilbratt \cite{pilbratt10}) and 
its PACS (Poglitsch et al. \cite{poglitsch10}) and SPIRE (Griffin et al. 
\cite{griffin10}) instruments, this 
has changed dramatically. With its 3.5~m passively cooled mirror it provides
the much improved spatial resolution (thus reduced source confusion) and the
 sensitivity needed for the next significant step in far-infrared studies of 
galaxy evolution. Members of the PACS instrument consortium, the Herschel 
Science Centre and mission scientist M. Harwit have joined 
forces in the {\em PACS Evolutionary Probe (PEP)} deep extragalactic survey,
to make use of this opportunity. PEP aims to resolve the cosmic 
infrared background and determine the nature of its constituents,
determine the cosmic evolution of dusty star formation and of the 
infrared luminosity function, elucidate the relation of far-infrared 
emission and environment and determine clustering properties.
Other main goals include study of AGN/host coevolution, and determination 
of the infrared emission and energetics of known high redshift galaxy 
populations.
  
\begin{figure}
\centering
\includegraphics[width=\columnwidth]{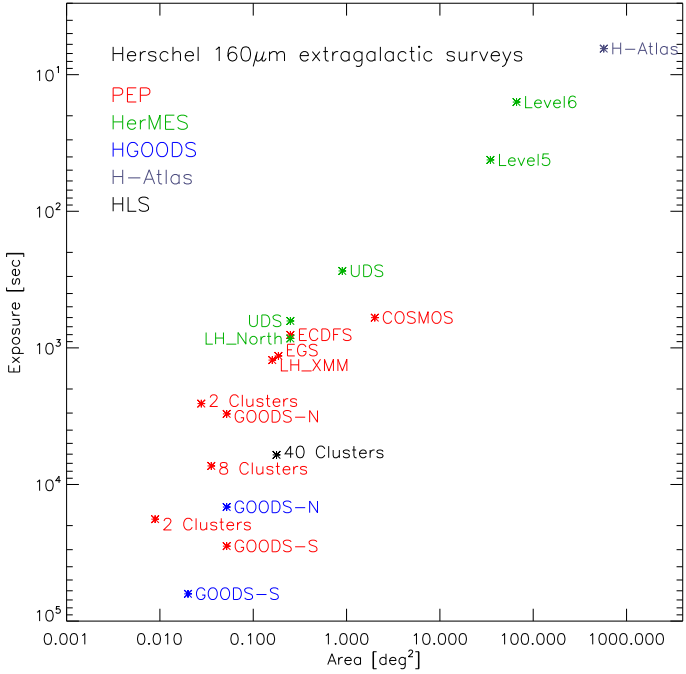}  
\caption{Area and 160$\mu$m exposure of PEP surveys (red) compared to other
major Herschel extragalactic surveys. Exposure is defined as survey time
(including overheads) multiplied with the ratio of PACS array size and 
survey area.
1000~sec exposure corresponds to an approximate 3$\sigma$ depth of 8~mJy.
Scaling this as exposure$^{-0.5}$ to smaller/deeper fields will be
too optimistic because of increasing overhead fractions and beginning source 
confusion.}
\label{fig:surveys}
\end{figure}

PEP encompasses deep observations of blank fields and lensing clusters, 
close to the \herschel\ confusion limit,
in order to probe down to representative high redshift galaxies, rather
than being restricted to individually interesting extremely luminous cases. 
PEP is 
focussed on PACS 70, 100, and 160~$\mu$m observations. SPIRE observations 
of the PEP fields are obtained in coordination with PEP by the HerMES survey 
(Oliver et al. \cite{oliver11}). Larger and shallower fields are observed by 
HerMES (70~deg$^2$) as well as by the H-ATLAS survey (570~deg$^2$, 
Eales et al. \cite{eales10}), while the GOODS-Herschel program
(Elbaz et al. \cite{elbaz11}) provides deeper observation in (part of) the 
GOODS fields that are also covered by PEP. Finally, the Herschel lensing 
survey (Egami et al. \cite{egami10})
substantially increases the number of lensing clusters observed with Herschel,
adding about 40 clusters to  the 10 objects covered by PEP.
Fig.~\ref{fig:surveys} compares for 160$\mu$m wavelength the area and 
exposure
of the PEP surveys (Table~\ref{tab:fields}) with that of these other major 
Herschel extragalactic surveys.

In this paper, we describe the field selection, observing strategy and
data analysis methods of PEP. We give a complete overview of the planned 
PEP observations and their execution status as of June 2011 
(Table~\ref{tab:fields}). We provide a detailed account of the science
demonstration phase (SDP) data sets for GOODS-N and Abell 2218, and give an 
overview of first science 
results.
%Throughout this paper,  we adopt an $\Omega_m =0.3$, $\Omega_\Lambda =0.7$ and $H_0=70$ km\,s$^{-1}$\,Mpc$^{-1}$ cosmology.

\section{Field selection}
A key element in selecting a field for a deep Herschel extragalactic survey
is the availability of a strong multi-wavelength database from X-rays to 
radio wavelengths, which is 
fundamental to many of the science results discussed in 
Sect.~\ref{sect:science}. In particular, deep optical, near-IR, and 
Spitzer imaging, as well as a comprehensive
set of photometric and spectroscopic redshifts are an essential asset for 
most of the far-infrared studies of galaxy evolution that we envisage. 
Deep X-ray data are invaluable for using the potential of
Herschel for studying the AGN -- host galaxy coevolution.

Another requirement is a low galactic far-infrared background
in order to minimize contamination by galactic `cirrus' structure and 
by individual 
galactic foreground objects. This naturally coincides with the selection
criteria of extragalactic surveys at other wavelengths. In the X-ray regime, 
for example, these are pushing
for a low galactic foreground obscuration. Given that
the power spectra of far-infrared
emission of galactic cirrus are steeply decreasing towards higher spatial 
frequencies (e.g., Kiss et al. \cite{kiss01}), such 
cirrus contamination is much less of a 
practical worry for point source detection in deep \herschel\ fields 
compared to the 
previous smaller cryogenic space telescopes. In any case, our blank field 
selection includes some of the lowest galactic foreground fields 
(Lockman Hole, CDFS, 100~$\mu$m sky brightness $\sim$0.4~MJy/sr), reaching 
up to $\sim$0.9~MJy/sr (COSMOS). Our individual cluster fields typically show a
100~$\mu$m galactic foreground of 1-2~MJy/sr,
with a maximum of 4.5~MJy/sr (Abell~2390).

\begin{table*}
\caption{AOR parameters used for the PEP fields}           
\centering                       
\begin{tabular}{l|rrrrrrr|rrrrrrr}      
\hline\hline              
Field        &\multicolumn{7}{c}{Nominal scan direction}&\multicolumn{7}{c}{Orthogonal scan direction}\\
   &Leg&Step&N$_L$&Angle& &N$_{Rep}$&N$_{AOR}$&Leg&Step&N$_L$&Angle& &N$_{Rep}$&N$_{AOR}$\\
   &\arcmin&\arcsec&&\degr&&&&\arcmin&\arcsec&&\degr&&&\\    
\hline                       
COSMOS          &85&Hom&Sq&   70&Arr& 1&24&85&Hom&Sq&  160&Arr& 2&25\\
Lockman Hole XMM&24& 50&30&    0&Sky& 2&10&24& 50&30&   90&Sky& 2&10\\
EGS             &67& 50&13&4 0.5&Sky& 2&13&10& 50&81&130.5&Sky& 2& 8\\
ECDFS           &30& 50&37&    0&Sky& 2& 8&30& 50&37&   90&Sky& 2& 8\\
GOODS-S 70/160  &17& 25&27&348.7&Sky& 2&48&11& 25&41& 78.7&Sky& 2&48\\
GOODS-S 100/160 &17& 25&27&348.7&Sky& 2&54&11& 25&41& 78.7&Sky& 2&54\\
GOODS-N         &17& 25&27&   41&Sky& 2&11&11& 25&41&  131&Sky& 2&11\\
Cl0024+16       & 6& 20&19&   45&Arr&15& 1& 6& 20&19&  315&Arr&15& 1\\
Abell 370       & 4& 20&13&   45&Arr&22& 1& 4& 20&13&  315&Arr&22& 1\\
MS0451.6-0305   & 4& 20&13&   45&Arr&22& 1& 4& 20&13&  315&Arr&22& 1\\
Abell 1689      & 4& 20&13&   45&Arr&18& 3& 4& 20&13&  315&Arr&18& 3\\
RXJ1347.5-1145  & 4& 20&13&   45&Arr&22& 1& 4& 20&13&  315&Arr&22& 1\\
MS1358.4+6245   & 4& 20&13&   45&Arr&22& 1& 4& 20&13&  315&Arr&22& 1\\
Abell 1835      & 4& 20&13&   45&Arr&22& 1& 4& 20&13&  315&Arr&22& 1\\
Abell 2218      & 4& 20&13&   45&Arr&14& 3& 4& 20&13&  315&Arr&14& 3\\
Abell 2219      & 4& 20&13&   45&Arr&22& 1& 4& 20&13&  315&Arr&22& 1\\
Abell 2390      & 4& 20&13&   45&Arr&22& 1& 4& 20&13&  315&Arr&22& 1\\
RXJ0152.7-1357  &10& 25&13&   45&Sky& 6& 3& 5& 25&25&  135&Sky& 2& 3\\
MS1054.4-0321   &10& 25&13&    0&Sky& 6& 3& 5& 25&25&   90&Sky& 2& 3\\
\hline                               
\end{tabular}
\tablefoot{For both the nominal and the orthogonal scan direction,
the table lists the scan leg length, the cross-scan separation step, the
number of scan legs, the scan angle and its reference system
(Sky or Array), the number of scan repetitions within an AOR and
the number of independent AORs obtained with those settings. 'Hom Sq' indicates
that the cross-scan separation and number of scan legs are defined
automatically by the AOT logic to produce a square map with `homogeneous'
coverage.}
\label{tab:aor}     
\end{table*}

Our largest field is the contiguous 2 square degree COSMOS (Scoville et al.
\cite{scoville07}) observed for about 200~hours to a 3$\sigma$ depth at 
160~$\mu$m of 10.2~mJy. At this level, integral number counts reach one source 
per 24~beams (Berta et al. \cite{berta10,berta11}), 
similar to the 5$\sigma$ 40~beams/source definition of the 
`confusion limit' used by, e.g., Rowan-Robinson et al. (\cite{roro01}). 
Slightly deeper observations have been obtained for the Lockman Hole 
(e.g. Hasinger et al. \cite{hasinger01}), the 
Extended Groth Strip EGS covered by the Aegis survey 
(Davis et al. \cite{davis07}), and the
extended Chandra Deep Field South (ECDFS) for which the name has been
coined in the X-rays  (Lehmer et al. \cite{lehmer05}) but a multitude of 
data exist at other wavelengths. Refined analysis of source confusion (Dole
et al. \cite{dole04}) suggests that with the 100~$\mu$m and 160~$\mu$m number 
counts turning over at a depth that is reached with PEP
(Berta et al. \cite{berta10,berta11}), deeper observations can still be 
extracted reliably, in particular when using position priors from very 
deep 24~$\mu$m data (e.g. Magnelli et al. \cite{magnelli09}). We make use of 
this in our deepest blank field observations 
which are centered on the GOODS fields (Dickinson et al. in
prep.), with
strong emphasis on the GOODS-S. GOODS-S also is the only field which we 
additionally
observe at 70~$\mu$m, to a depth where \spitzer\//MIPS would be confused.
For redshifts around 1, our blank field observations sample a range of 
environments from the field to moderately massive clusters, which are 
known in  particular in COSMOS due to its large angular size but still
excellent multiwavelength characterisation. In order
to extend this to the full range of environments at this redshift, we 
add dedicated observations of two of the best studied massive z$\sim$1 
clusters, RXJ0152.7-1357 (e.g. Ebeling et al. \cite{ebeling00}) 
and MS1054.4-0321 (e.g. Tran et al. \cite{tran99}).

We use the amplification provided by massive galaxy clusters to study
sources that would otherwise be too faint for direct observations even in our
deepest blank fields, and to provide highest quality SEDs on more luminous
objects that are also detectable in the blank fields. For that purpose, we 
have selected ten of the best studied lensing 
clusters at redshifts z$\sim$0.2--0.5. With 4\arcmin\ size our observations 
for these fields are optimized for studying lensed background objects. 
They will 
additionally detect part of the cluster members from the central region, 
but not the cluster outskirts and the adjacent field population.   

Main parameters of our survey fields are summarized in Table~\ref{tab:fields}.

\section{Observing strategy}
\label{sect:obsstrat}
The scan map is the PACS photometer observing mode that is best suited 
to get deep maps close to the confusion limit, for regions that are 
dominated by (almost) point sources. During a PACS prime mode scan map 
observation, the telescope moves back and forth in a pattern of parallel
scan lines that are connected by short turnaround loops. During such a scan, 
the  PACS photometer arrays take data samples at 10~Hz frequency.
Given the presence of 1/f noise in the PACS bolometers, we consistently
follow the recommendation to adopt a medium scan speed of 20~arcsec/sec
that reduces the effects of 1/f noise on point source sensitivity compared 
to slow scan speed of 10~arcsec/sec, while
not yet causing point spread function (PSF) degradation which is present 
in fast scans at 
60~arcsec/sec because of data sampling and detector time constants. 

Much of the science from deep far-infrared surveys strongly benefits
from the availability of ancillary multiwavelength data that historically
have been obtained in fields of a certain shape and orientation. We have
matched our blank field data to such constraints (e.g. the shape/orientation
of the mid-IR GOODS fields) by typically defining scan layouts in
`sky' coordinates rather than `array' coordinates that rotate with epoch of 
observation. This is also facilitating the combination of data from different 
epochs. Exceptions to this approach are the COSMOS field (discussed below) 
and the small lensing
cluster fields for which there is no preferred orientation.

Reaching the desired depths for our fields requires multiple passages
over a given point in the sky. Rather than simply repeating a single scan
we use this to improve the redundancy of the data. Specifically, the
observation setup for PEP includes
\begin{itemize}
\item Scans in both nominal and orthogonal directions. While the 
highpass-filtered data reduction discussed in 
Sect.~\ref{sect:dataanalysis} does
not require this, such crosslinking is essential for alternative reductions
with full inversion algorithms such as the MadMap implementation in the 
Herschel Common Science System (HCSS)\footnote{HCSS is a joint 
development by the Herschel Science Ground 
Segment Consortium, consisting of ESA, the NASA Herschel Science Center,
and the HIFI, PACS and SPIRE consortia} software environment.
Typically, we obtain the same number of astronomical observing requests
(AORs) in nominal and orthogonal
scan direction, often concatenated in pairs. For the very elongated EGS
and z$\sim$1 clusters we overweight maps scanning along the long axis, 
in order to reduce overhead losses that are caused by the time spent
in scan turnaround loops. 
\item Small cross-scan separations, specifically the size of one of the 
eight PACS `blue' detector array matrices ($\sim$50\arcsec) or fractions 
1/2 or 1/2.5 of it. Simple models 
demonstrate that for such a scan pattern homogeneous coverage maps are 
produced already from a 
single detector matrix, for any relative orientation of the scan 
direction and the 
inter-matrix gaps. By definition, this pattern also averages out sensitivity
variations between detector matrices over most of the final map.
If this redundant mapping scheme led to short execution times of a single
scanmap over the field but deeper observations were needed, the scanmap 
was repeated within an AOR to reach a total execution time between one and 
a few hours.  
\item Often, many AOR pairs with a plausible execution time that is not 
exceeding a few hours each, are still 
needed to achieve the required depth. Then, AOR positions may be dithered by a 
fraction of the cross-scan separation to further improve spatial redundancy, 
and the corner of the map where the scan is started may be varied. 
\end{itemize}

\begin{figure}
\centering
\includegraphics[width=\columnwidth]{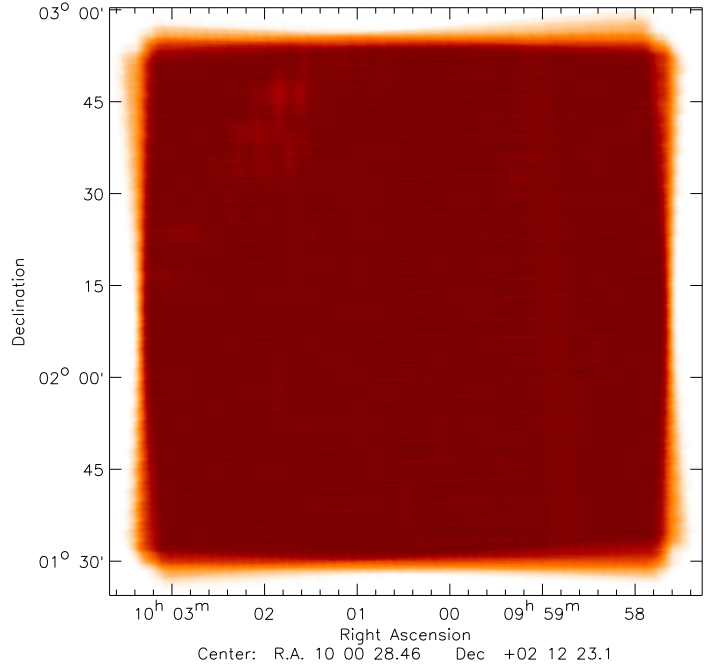}  
\caption{Actual 160~$\mu$m coverage map of the COSMOS field. Spots
of reduced coverage near the top left of the map are due to dropping of data
 because of `speed bumps' (see text).}
\label{fig:cosmoscoverage}
\end{figure}

The two square degree COSMOS field is a special case where the described
observing strategy would lead to extremely long individual AORs. During
one of the about two month long \herschel\ visibility periods of COSMOS, 
the PACS arrays are always similarly oriented on this region of the sky, 
with a position angle of the long axis about 20$\pm10$~\degr\/. This 
permits an AOR setup in the efficient
`homogeneous coverage' mode in array coordinates, keeping the individual 
AOR length below 5~hours but staying well matched to the roughly square
nonrotated orientation of many COSMOS ancillary data sets. This is
illustrated by the actual coverage map obtained with PACS 
(Fig.~\ref{fig:cosmoscoverage}).  

Table~\ref{tab:aor} provides the key parameters of the actual PACS AOR 
implementations for each field. No significant source variability is expected
for the dust-dominated emission of almost all detected sources. For this
reason no timing contraints needed to be applied in the scheduling. For
practical reasons scheduling of all AORs of a field during a visibility period
was aimed for, and was typically but not always achieved. For fields near the
plane of the ecliptic (COSMOS), asteroid passages may introduce another 
time dependent 
factor, as clearly demonstrated in mid-infrared detections during 
\spitzer\ observations of the COSMOS field (Sanders et al.
\cite{sanders07}). The contrast between galaxies and asteroids is more
favourable in the far-infrared. Still, bright asteroids would be
detectable in individual maps if present but were not identified when 
differencing our individual COSMOS maps from a coaddition.    

SPIRE maps for most of the PEP fields are obtained in coordinated observations 
by the HerMES key program (Oliver et al. \cite{oliver11}). For the 
two z$\sim$1 clusters we implemented within PEP simple
10\arcmin$\times$10\arcmin\ `large' SPIRE scanmaps in nominal scan speed, 
spatially dithering between five concatenated independent repetitions.

\section{Data Analysis}
\label{sect:dataanalysis}
\subsection{Reduction of scanmap data and map creation}

For scanning instruments with detectors that have a significant
1/f low frequency noise component,  map creation usually follows one 
of two alternative routes. 
One is using full `inversion' algorithms as widely applied by the cosmic 
microwave background community and the other uses highpass filtering of 
the detector timelines and subsequent direct projection, frequently
used for \spitzer\ MIPS 70 or 160~$\mu$m reductions. An algorithm of the 
first `inversion' type is
available in the HCSS \herschel\ data processing in the form of an 
implementation and adaption to Herschel of a version of the MadMap code
(Cantalupo et al. \cite{cantalupo10}). The alternative option that we adopt
and describe in more detail below is using highpass filtering of the 
detector timelines and a direct `naive' mapmaking. 
This choice is made because for our particular case of deep
field observations, MadMap presently does not reach the same point source 
sensitivity, and the preservation of diffuse emission is not important for our
science case. As noted, the cross-linked design of the PEP observations 
however does permit the future application of such inversion codes.

\begin{figure}
\centering
\includegraphics[width=\columnwidth]{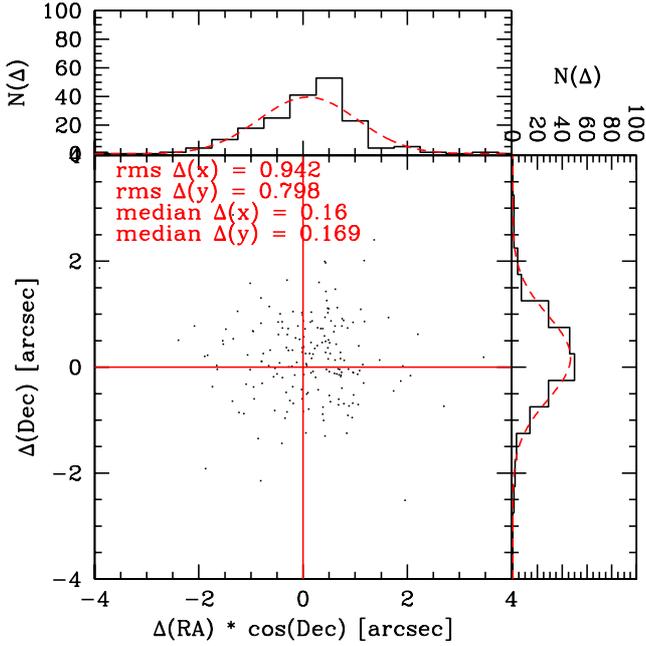}  
\caption{Offsets of GOODS-N 100~$\mu$m sources detected at $>5\sigma$
in a `blind' extraction, compared to the corresponding 24~$\mu$m positions 
of the MIPS catalog used by Magnelli et al. (\cite{magnelli09}). 
A global offset  has already been subtracted when deriving the PACS map.}
\label{fig:blindpos}
\end{figure}

Our reduction first proceeds on a per AOR level and is based on scanmap 
scripts for the PACS photometer pipeline (Wieprecht et al. \cite{wieprecht09})
in HCSS, with parameter settings and additions
optimized for our science case. After retrieving PACS data and satellite
pointing information we apply the first reduction steps to the 
time-ordered PACS data 
frames, identifying functional blocks in the data, flagging bad pixels,
flagging any saturated data, converting detector signals from digital units
to volts and the chopper position from digital units to physical angle.

After adding to the time-ordered data frames the instantaneous pointing
obtained from the Herschel pointing product, we apply `recentering' 
corrections. These are derived by comparing PACS maps obtained in a separate 
first processing of partial datablocks to deep 24~$\mu$m catalogs with accurate 
astrometry.
Deep radio catalogs can also be used successfully. The corrections
are derived by stacking at the positions of 24~$\mu$m sources a PACS 
100~$\mu$m or 70~$\mu$m map that is obtained from typically 
about 15 minutes of data, and measuring 
the offset of the stacked PACS detection. The measured position offset is then
corrected for this block of data in the original frames. The blocks used 
were typically
restricted to one scan direction (all `odd' or `even' scanlegs of a map 
repetition, or subsets for the large COSMOS maps). By this procedure we correct
for (1) the global offset pointing error of \herschel\ which is of order 
2\arcsec\ RMS and thus nonnegligible compared to PACS beamsizes, (2) 
differential pointing errors between different AORs that would lead to PSF 
smearing, in particular if AORs have been observed spread over long periods 
with different spacecraft orientations, (3) residual timing offsets
between PACS data and pointing information that manifest themselves in
small offsets between odd and even numbered scanlines and (4) pointing 
drifts over $\gtrsim$0.5\,hour timescales
within an AOR. Overall this leads to subarcsecond astrometry and reduces
unnecessary PSF smearing by pointing effects. As a verification of this
approach, Figure~\ref{fig:blindpos} compares the positions from a {\em blind}
100~$\mu$m catalog based on the final GOODS-N map with 24~$\mu$m positions.

Short glitches in the detector timelines caused by ionizing particle hits  
are flagged and interpolated with an HCSS implementation of the 
multi-resolution median transform, developed by Starck \& Murtagh 
(\cite{starck98}) to 
detect faint sources in ISOCAM data. The different signatures of real sources 
and glitches in the pixel timeline at medium scan speed are discriminated 
by a multi-scale 
transform. This procedure is known to produce false detections on bright 
point sources. This was no problem for our deep field data except for a 
few bright COSMOS
sources for which we locally reduced the glitch detection sensitivity.

\begin{figure}
\centering
\includegraphics[width=0.7\columnwidth]{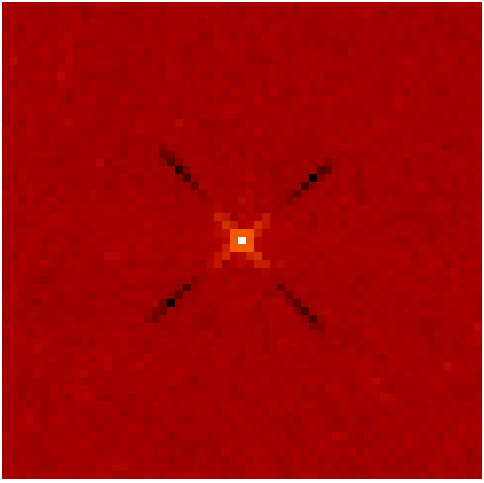}  
\caption{Correlation map extracted from GOODS-N 100~$\mu$m data, reflecting
correlations induced by map projection and -- in scan direction -- by 
residual low frequency noise and highpass filtering. Correlations 
across the GOODS-N map are highly uniform and depend only on the 
relative $\Delta$x, $\Delta$y pixel positions, represented here with
respect to the central pixel.
}
\label{fig:corrmap}
\end{figure}

The `1/f' noise of the PACS photometers, in fact roughly 
$\propto f^{-0.5}$ over the relevant frequencies, is removed
by subtracting from each timeline the timeline filtered by a running box 
median filter of radius 15 samples (30\arcsec) at 70~$\mu$m or 100~$\mu$m and 26 
samples (52\arcsec) at 160~$\mu$m. A mask is used to exclude sources from the 
median derivation. The mask is created by thresholding a S/N
map produced from a smoothed coadded map that is including all AORs. Tests 
were done adding simulated sources to the timelines before 
masking and before highpass filtering. These tests were done on the 
real timelines of the full Lockman Hole AOR set for the red filter, thus 
implicitly including all real sky structures. They indicate
that the filtering modifies the fluxes by less than 5\% for masked point 
sources and $\sim$16\% for very faint unmasked point sources. These results
apply to `extragalactic deep field' skies that are composed of many point 
sources and to our specific highpass filter setting. Extended sources need
different filtering radii or reduction methods.

Flatfielding and flux calibration is done via the standard PACS pipeline
calibration files. Because of the excellent in-flight stability of
the photometer response, no use is made of the observations
of the calibration sources internal to PACS that are obtained within each AOR. 
PEP science demonstration
phase data were reduced using earlier versions of the responsivity calibration
file and corrections derived on the stars $\gamma$Dra, $\alpha$Tau, and
$\alpha$CMA; they are within 5\% of the currently valid Version 5 of
the reponsivity calibration file that has been validated over a wider
set of flux calibration stars and asteroids.  

\begin{figure*}
\centering
\includegraphics[width=9.cm]{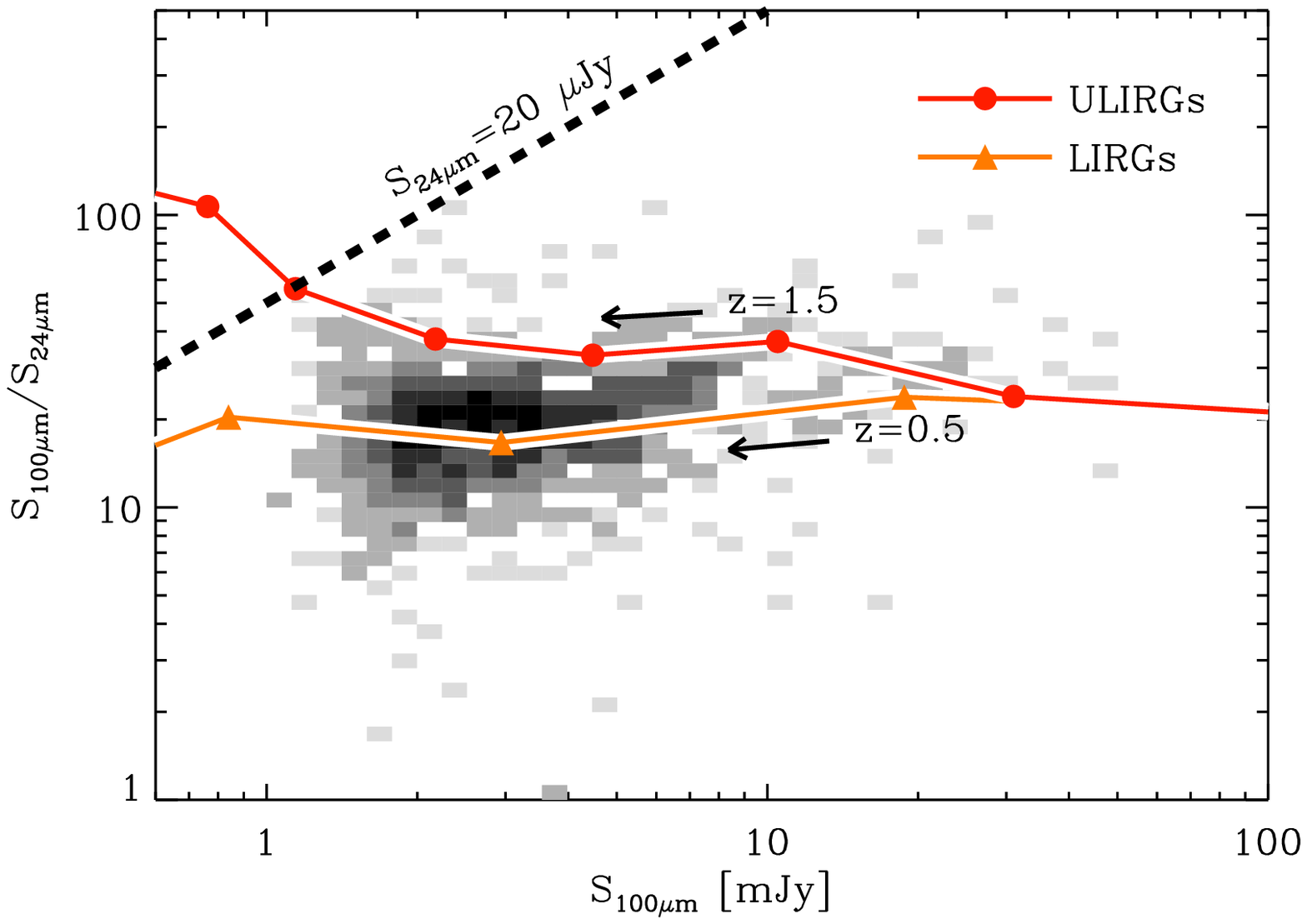}
\includegraphics[width=9.cm]{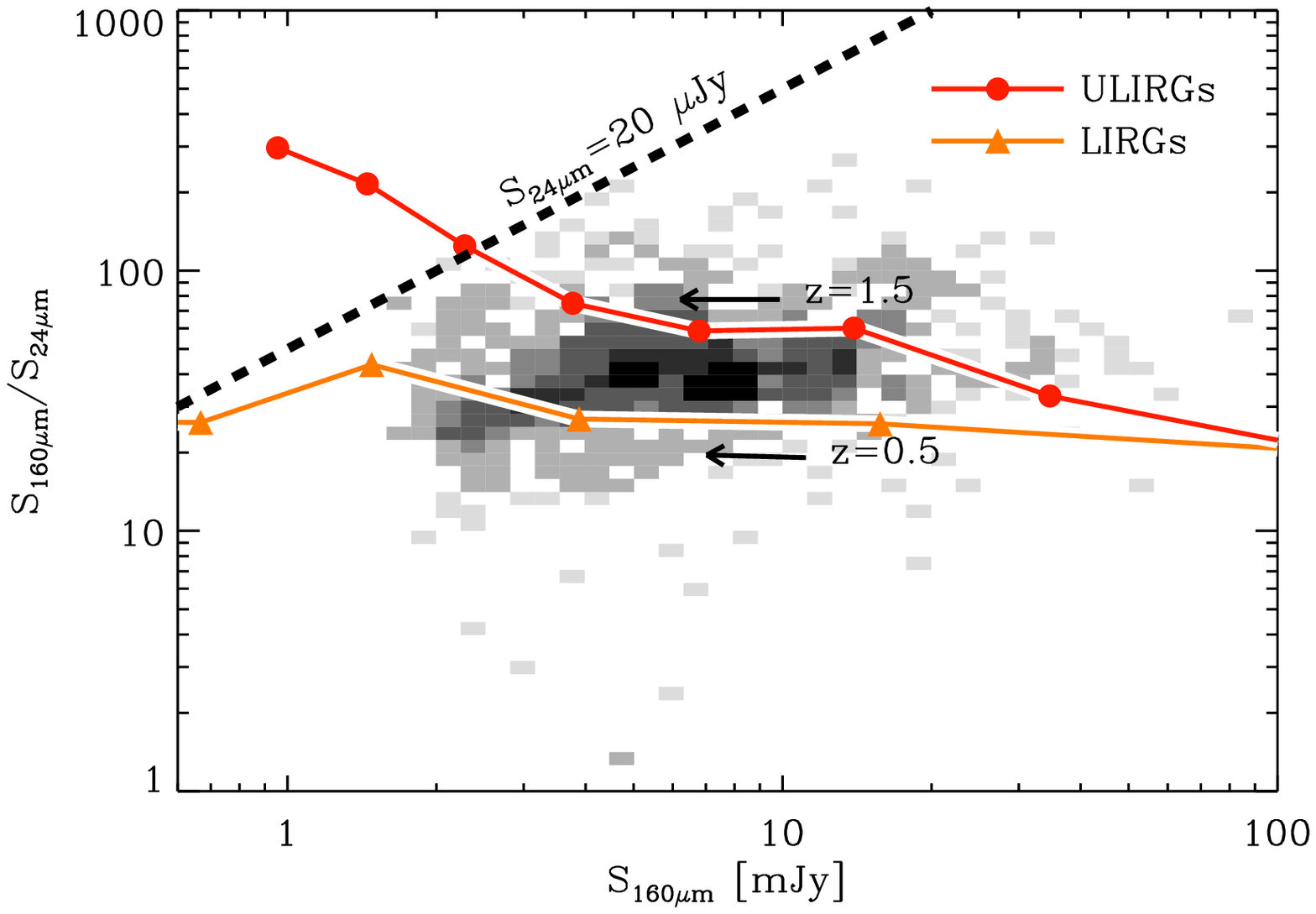}
\caption{\label{fig: MIPS PACS}\small{
\textit{(Left)} PACS 100 $\mu$m to MIPS 24 $\mu$m flux density ratio as a function of the PACS 100 $\mu$m prior catalog flux density. Shaded grey regions show the space density distribution of galaxies observed in the GOODS-S field. In that field, the MIPS 24 $\mu$m and PACS 100 $\mu$m catalogs reach at 3-$\sigma$ limit of 20~$\mu$Jy and 1.2 mJy, respectively. Red dots and orange triangles present the evolution with redshift of the PACS-to-MIPS flux density ratio of ultra-luminous (ULIRGs) and luminous (LIRGs) infrared galaxies, respectively, predicted using the Chary \& Elbaz (\cite{chary01}) library. Each symbol corresponds to a given redshift, in intervals of $\Delta z=0.5$. On each track, we indicate on one point its corresponding redshift as well as as the path followed for increasing redshifts. The dashed black line represents the limit of the parameter space reachable using the MIPS 24 $\mu$m catalog available in this field; all the PACS 100 $\mu$m sources below this line would have a MIPS 24 $\mu$m counterpart in this catalog. \textit{(Right)}  PACS 160 $\mu$m to MIPS 24 $\mu$m flux density ratio as
function of the PACS 160 $\mu$m flux density. Symbols and lines are the same as in the \textit{left} panel.}}
\end{figure*}

\begin{figure}
\centering
\includegraphics[width=\columnwidth]
{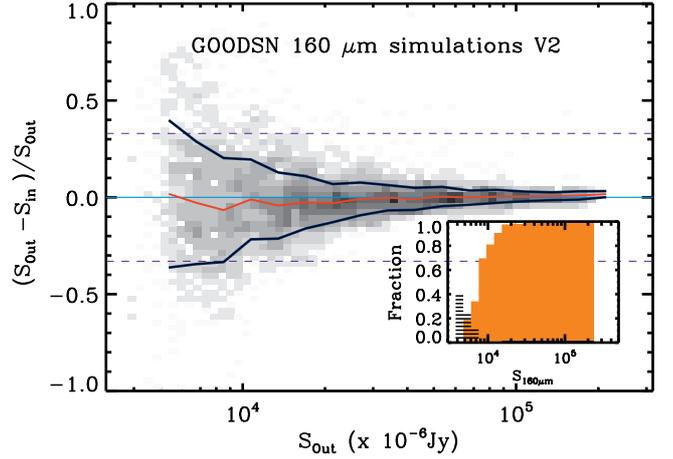}  
\includegraphics[width=\columnwidth]
{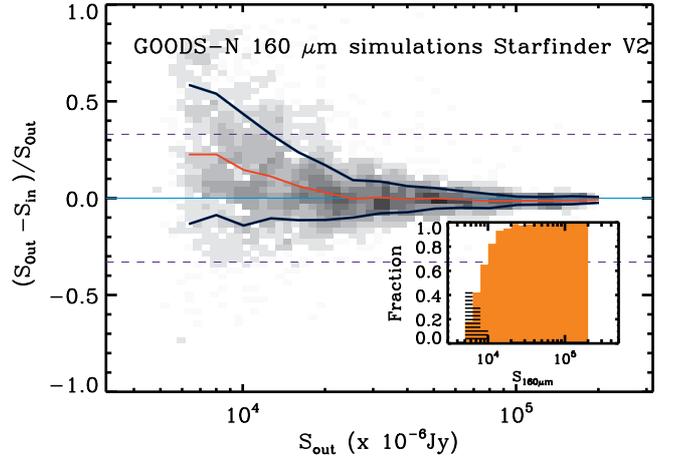}  
\caption{Results of artifical source addition experiments for  the 
GOODS-N 160~$\mu$m extractions. Top: prior extraction, Bottom: blind extraction.
The red lines represent the average photometric accuracy, blue lines set 
the standard deviation observed in
each flux bin (after 3$\sigma$ clipping). The systematic boosting
in the blind extraction is corrected in the final blind catalog.
Given that, input and output fluxes are
consistent with each
other within a few percent, testifying that PACS fluxes are reliably
extracted and retrieved by the adopted source extraction methods.
The orange histogram represents the detection rate (or completeness) 
computed on the artificial injected sources, while black histograms denote 
cp the fraction of spurious sources as a function of output flux.
}
\label{fig:sims}
\end{figure}

\begin{figure}
\centering
\includegraphics[width=\columnwidth]{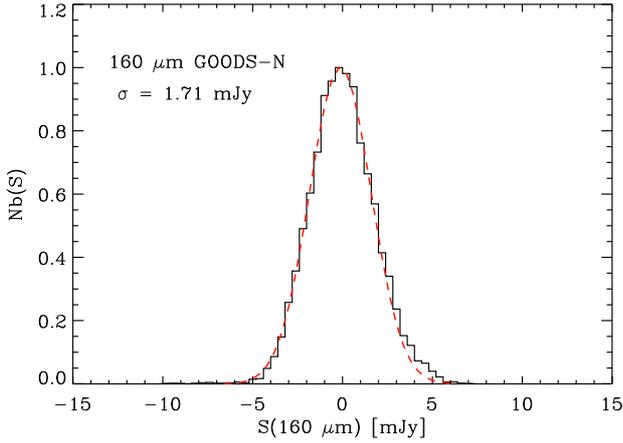}  
\caption{Noise estimate from randomly placing apertures on the
GOODS-N 160~$\mu$m residual map.}
\label{fig:goodsnnoisehisto}
\end{figure}

Before creating maps from the timelines we discard various unsuitable
parts of the data:
\begin{enumerate}
\item Observations of the internal calibration sources and a short subsequent
interval during which the signal re-stabilizes. SDP observations
had an erroneously large frequency of such calibration blocks.  
\item Data taken during the turnaround loops between scans. The nominal 
pointing accuracy is not guaranteed during this phase, and highpass filtering
effects on source flux are more severe if the telescope slows down strongly 
or even stops during a subtantial part of the filtering window. 
\item Before a reduction of the star tracker operating temperature in 
\herschel\ operational day 320, `speed bumps' 
in the satellite movement occured, when stars crossed irregularly behaving 
star tracker pixels. This led to an unkown mispointing. These events can be 
identified by comparing the two angular velocities obtained from 
(1) differentiating the positions on sky reported in subsequent 
entries of the pointing product and 
(2) the direct angular rate information in the pointing product.
\item Blue PACS bolometer data occasionally show fringes in the signal
due to magnetic field interference that is heavily aliased into the timelines
and final maps. Maps created separately for each individual scan line
were visually inspected and scans clearly affected by fringing excluded
from final mapping.
\end{enumerate}

Maps are obtained from the timelines for each AOR via the HCSS `photProject'
projection algorithm, which is equivalent to a simplified version of the 
`drizzle' method
(Fruchter \& Hook \cite{fruchter02}). Given the high data redundancy in the
deep fields, PSF widths and noise correlation in the final map
can be reduced choosing smaller projection drops than the physical PACS pixel
size. Drop sizes between 1/8 and 1/4 of the physical PACS pixel size
were used, depending on redundancy of the individual maps for each field. 
Weights of the different detectors in the projection consider the inverse
variance derived from the noise in the dataset itself.

Maps from each AOR were coadded into final maps, weighting the indivudal
maps by the effective 
exposure of each pixel. The final error
map was computed as the standard deviation of the weighted
mean. PSF fitting using the methods described in Sect.~\ref{sect:extract} 
assumes errors that are uncorrelated between
neighbouring pixels. In practice, correlations exist
due to projection and due to to the correlations that are 
caused by residual
1/f noise in the filtered timelines, in particular along the scan 
direction. We have verified that, because of the high redundancy of 
the data, these
correlations are close to uniform across the final map, with less than 
2\% variation on the correction factor that is derived below. Thus we
derived from PSF shape and correlation information a mean correlation 
correction factor which was then accounted
for in the errors on the extracted fluxes.
A correlation map is constructed, starting by collecting series of paired pixel
values with same relative pixel coordinate offsets i,j. We take
values $\Delta$f defined as the deviation in flux of a pixel in an individual
AOR map from the corresponding pixel in 
the final map. This is a deviation from the mean with expectation value 0. 
Such values are taken for a large number of pairs in 
different positions in each AOR map and from different AORs. For each 
pixel-pairs series  $\Delta$f$_{1,2}$, which corresponds to a specific i,j 
pixel offset, a correlation coefficient is calculated: 
\begin{equation}
\rho(i,j)=\frac{\sum{\Delta f_1 \Delta f_2}}{\sqrt{\sum{(\Delta f_1)^2}}
\sqrt{\sum{(\Delta f_2)^2}}}
\end{equation}
The correlation coefficients are 
stored as a map and written to the i,j position relative 
to the central pixel. Figure~\ref{fig:corrmap} shows an example correlation 
map. Normal error propagation for a weighted sum 
$g(x_1..x_n) = \sum_{k=1}^n{a_k \cdot x_k}$, where the error are $\sigma_k$ 
and correlation coefficients $\rho(k,l)$ is:
\begin{equation}
 \sigma_g^2 = \displaystyle\sum_{k,l=1}^{n}{a_k \sigma_k \cdot a_l \sigma_l \cdot \rho(k,l)}
\end{equation} 
Given a PSF stamp with pixel values $P_k$ and a correlation between
every two pixels $\rho(i,j)$ known from their relative position, the
correlation correction factor to the derived errors is:
\begin{equation}
f^2 = \frac{\displaystyle\sum\limits_{k}{P_k^2}}{\displaystyle\sum
\limits_{i,j}{P_i P_j \rho(i,j)}}
\end{equation}
$f$ is the ratio of the propagated error without correlations and the error 
calculated with the correlation terms.
This assumes a near uniform error map on a scale of a PSF. 
For the typical pixel sizes 
(2\arcsec\ and 3\arcsec\ at 100~$\mu$m and 160~$\mu$m) and
projection parameters used in the
PEP SDP reductions, the corrections are about $f \approx 1.4$ for the 70~$\mu$m
and 100~$\mu$m maps and $f \approx 1.6$ for the 160~$\mu$m maps.

Before availability of version 6 of the ArrayInstrument calfile that
is containing the spatial transformation from PACS focal plane to sky,
160~$\mu$m data showed a $\approx$ 1\arcsec\ spatial 
offset from 70~$\mu$m or 100~$\mu$m
data. We corrected for this ad hoc in the 160~$\mu$m map fits headers, 
using offsets derived from a comparison of preliminary catalogs in the two 
bands.

\subsection{Catalog creation}
\label{sect:extract}
We extract source catalogs using point source fitting routines outside
HCSS. We have used both blind extraction via the Starfinder
PSF-fitting code (Diolaiti et al. \cite{diolaiti00}) and a guided extraction
using 24~$\mu$m source priors, following the method described in Magnelli
et al. (\cite{magnelli09}). We fitted with point spread functions extracted
from the maps. Since these observed PSFs are limited in radius, we used
for aperture corrections point spread functions obtained on 
Vesta\footnote{accompanying PACS ICC document PICC-ME-TN-033 version 0.3}. In 
order to match the observations, these were rotated
to match the satellite position angle for each observation of a field,
coadded
and slightly convolved with a gaussian to match the actual FWHM of the
combined map. For our reduction methods and projection into 2\arcsec\ map 
pixels at 70 and 100~$\mu$m and 3\arcsec\ at  160~$\mu$m, we have PSF FWHM for
the 70, 100, and 160~$\mu$m maps of 6.46\arcsec, 7.39$\pm$0.10\arcsec, and
11.29$\pm$0.1\arcsec\/. The FWHM values are 
from gaussian fits to the core of the observed PSF.
For the 100 and 160~$\mu$m widths we quote the error 
of the mean of measurements from five different fields.

All PEP blank fields benefit from extensive multi-wavelength coverage 
that is allowing
guided extraction based on source positions at shorter wavelengths,
where the depth and resolution of the observations are higher.
This approach resolves most of the blending issues encountered in dense
fields and allows straightforward multi-wavelength association (Magnelli
et al. \cite{magnelli09}, \cite{magnelli11}; 
Roseboom et al. \cite{roseboom10}). However, to use this powerful method,
prior source catalogs have to contain all the sources in the PACS images.
Deep MIPS 24~$\mu$m observations, available for all our blank fields, should
fulfill this criterion since they have higher resolution and are
deeper than our current PACS observations.
Moreover, since the 24~$\mu$m emission is strongly correlated
with the far-infrared emission, those catalogs will not contain a large excess
of sources without far-infrared counterparts.
This largely avoids deblending far-infrared sources into several
unrealistic counterparts, as could happen when using an optical
prior catalog with very high source density.
Figure \ref{fig: MIPS PACS} illustrates the validity of the MIPS 24 $\mu$m
observations as PACS prior source positions.

At 100 $\mu$m, models and observations predict typical PACS-to-MIPS flux ratios
in the range 5-50. A MIPS 24~$\mu$m catalog $50$ times deeper than the
PACS observations is thus suited to perform guided source extraction.
This is illustrated, for GOODS-S, in the \textit{left} panel of Figure
\ref{fig: MIPS PACS}, where the observed PACS population lies well-below
the boundary of the parameter space reachable by the MIPS 24 $\mu$m
catalog.
In all our blank fields, deep MIPS 24~$\mu$m catalogs fulfill this criterion
(i.e., GOODS-S/N, COSMOS, LH, ECDFS, EGS\dots).

\begin{figure*}
\centering
\includegraphics[width=\textwidth]{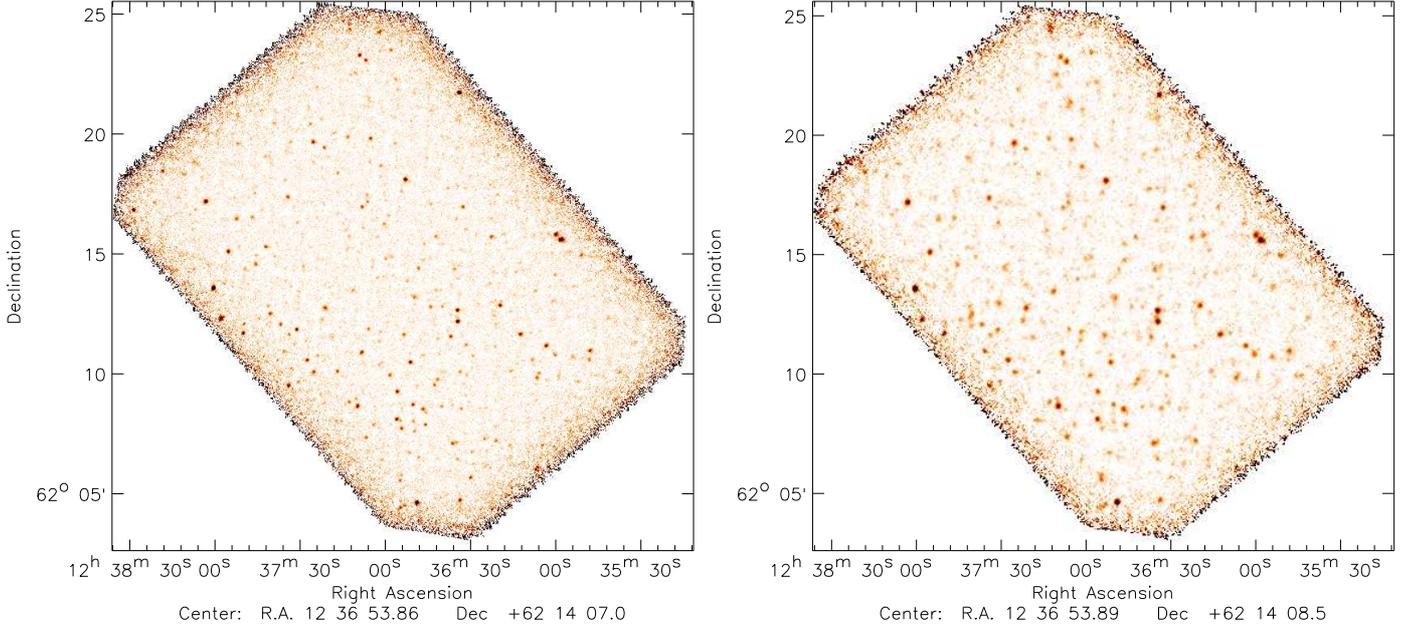}  
\caption{Science demonstration phase maps of the GOODS-N field. 
Left: 100~$\mu$m. Right: 160~$\mu$m}
\label{fig:goodsnsdp}
\end{figure*}

\begin{figure*}
\centering
\includegraphics[width=\textwidth]{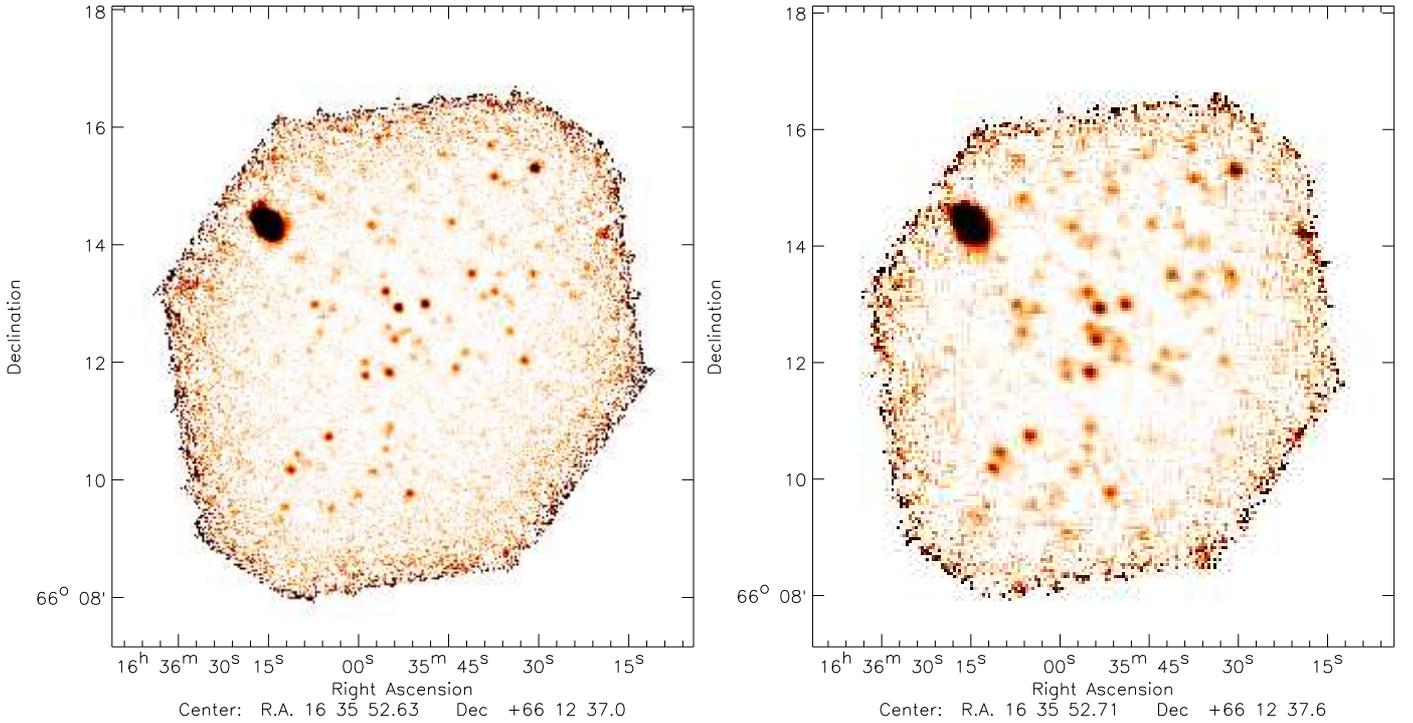}  
\caption{Science demonstration phase maps of the Abell 2218 lensing cluster
field. Left: 100~$\mu$m. Right: 160~$\mu$m}
\label{fig:a2218sdp}
\end{figure*}

At 160 $\mu$m, models and observations have typical PACS-to-MIPS flux ratios in
the range 15-150.
In all but one field, the deep MIPS 24~$\mu$m observations are at least
150 times deeper than our PACS observations; thus, they can be used as
prior source positions.
In GOODS-S, the deepest MIPS 24~$\mu$m observations are only 100 times
deeper than our PACS observations.
This limitation can be observed, in the \textit{right} panel of Figure
\ref{fig: MIPS PACS}, as a slight truncation, at faint 160~$\mu$m flux
density, of the high-end of the dispersion of the PACS-to-MIPS flux ratio.
This truncation will introduce, at faint flux density, incompleteness in
our prior catalog.
However, we also observe that even at this faint 160~$\mu$m flux density,
the bulk of the population has a PACS-to-MIPS flux ratio of $\thicksim60$.
The incompleteness introduced by the lack of MIPS
24 $\mu$m priors should thus be low or at least lower than the incompleteness
introduced by  source extraction methods at such low S/N.
This was checked by comparing the GOODS-S 160 $\mu$m catalog
obtained using blind source extraction with that obtained using guided
source extraction: we find no significant difference, at faint 160~$\mu$m
flux densities, in the number of sources in those two catalogs.
In line with these finding, Magdis et al. (in prep) find very low
fractions of 24$\mu$m undetected sources in very deep data from the 
GOODS-Herschel
key program, $<2\%$ for ratios of detection limits S(100)/S(24)$>$43 and
$<1\%$ for S(160)/S(24)$>$130.

Therefore, for all fields with deep MIPS 24~$\mu$m observations, we also
extract source catalogs with a PSF-fitting method using 24~$\mu$m source
positions as priors and  following the method described
in Magnelli et al. (\cite{magnelli09}).
We used the same PSFs and aperture corrections as for the blind
source extraction.
Blind and prior catalogs were compared to verify the consistency between
those two methods.

Completeness, fraction of spurious sources and flux reliability were
estimated by running Monte Carlo simulations. Up to 10000 artificial 
sources were added to PACS science maps, and then extracted with 
the same techniques and configurations adopted for real source 
extraction. In order to avoid crowding, many such frames were created, 
each including a limited number of artificial sources. The
number of frames and the number of sources added in each one depend on the
size of the field under analysis and range between 20 and 500 sources 
(GOODS fields or COSMOS) per frame, repeated up to reaching the 
total of 10000. These synthetic sources cover a large range in
flux, extending down to 0.5$\sigma$ ($\sigma$ being the measured rms noise
in the PACS maps). The flux distribution follows the detected number counts,
extrapolated to fainter level by means of the most successful fitting 
backward evolutionary model predictions
(see Berta et al. \cite{berta11}). Sources are modelled using the Vesta PSF,
manipulated as described above.

Figure~\ref{fig:sims} shows an example of results in the GOODS-N field at 
160~$\mu$m. Completeness is defined here as the fraction of sources 
that have been detected with a photometric accuracy of at least 50\% 
(Papovich et al. \cite{papovich04}). Spurious sources are
defined as those extracted above 3$\sigma$ with an input flux lower than
3$\sigma$(Image). The systematic flux boosting
in the blind extraction is corrected in the final blind catalog on the basis 
of these simulations.

Noise was estimated by extracting fluxes through 10000 apertures randomly
positioned on residual maps. Figure~\ref{fig:goodsnnoisehisto} 
shows the distribution of
the extracted fluxes, peaking around zero, as expected for a well subtracted
background, and showing a Gaussian profile.

\section{Science demonstration phase data}

Figures~\ref{fig:goodsnsdp} and \ref{fig:a2218sdp} show the 100~$\mu$m and
160~$\mu$m maps of the GOODS-N and Abell 2218 fields as obtained during the 
Herschel science demonstration phase. A conservative threshold of 90\% 
completeness is reached for the blind catalogs near 7~mJy and 15~mJy for 
100~$\mu$m and 
160~$\mu$m 
in the main parts of both GOODS-N and A2218. Above that level, the 
blind catalogs contain 153 and 126 sources for GOODS-N and 49 and 47 for 
Abell 2218, respectively.
At the time of these observations, the \herschel\ scan maps
were still exhibiting larger turnaround overheads than implemented later,
and the PACS scanmap AOR had an unnecessarily large frequency of internal 
calibrations. Both these factors do not significantly affect the maps 
which are based on highpass-filtered reductions. With the exception of
the overheads, these observations are representative
for the results achievable in later mission periods during the 
observing times listed in Table~\ref{tab:fields}, which reflect the
later reduced overheads.  

\begin{figure}
\centering
\includegraphics[width=\columnwidth]{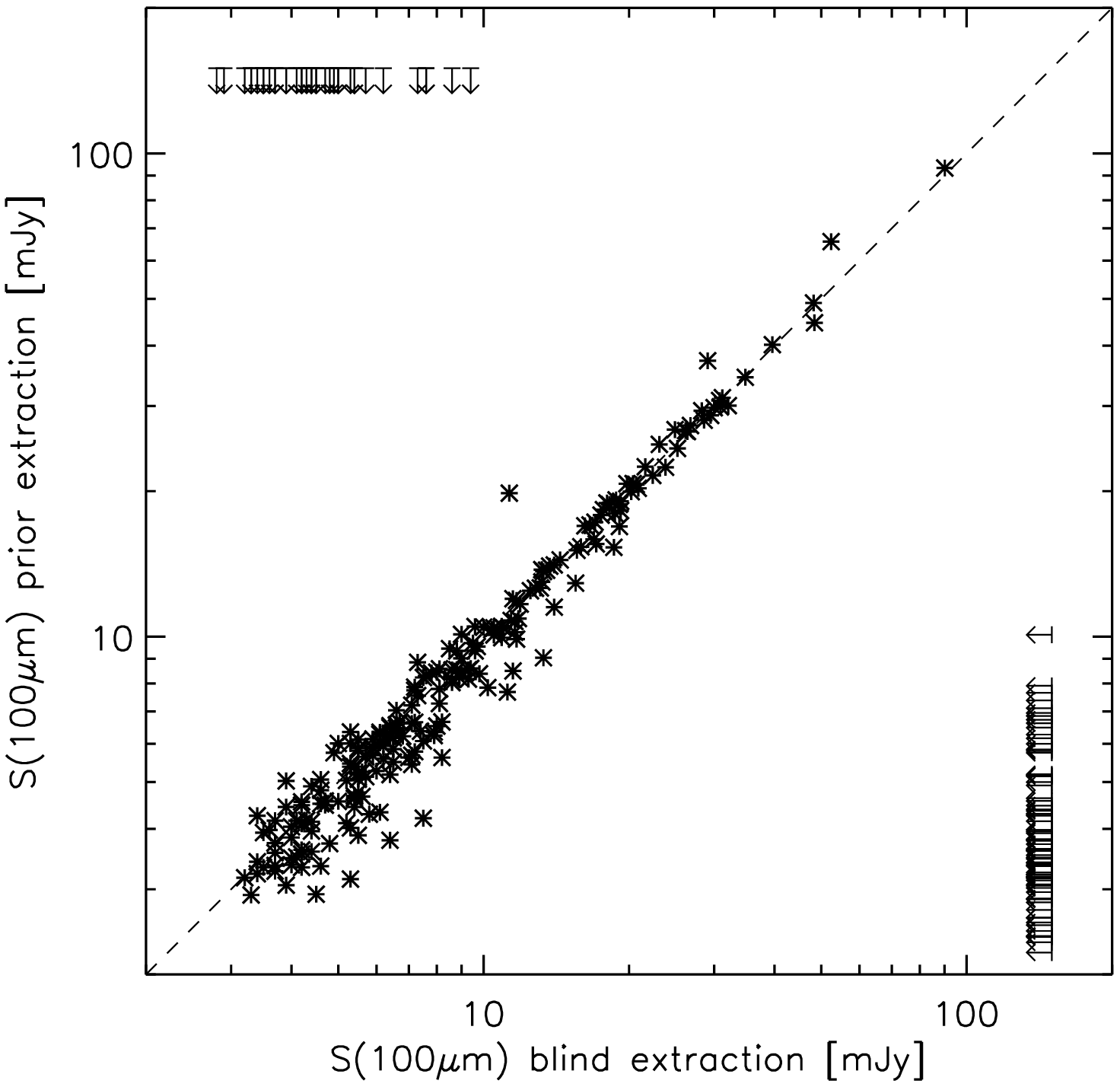}  
\caption{Comparison of blindly extracted and prior extracted 100~$\mu$m fluxes
for the GOODS-N field.}
\label{fig:blindprior}
\end{figure}

Figure~\ref{fig:blindprior} compares 100~$\mu$m fluxes for the GOODS-N
field between the blind (starfinder) extraction and extraction based
on 24~$\mu$m priors. The blind
detections have here been associated a posteriori to 24~$\mu$m sources,
using a modification of the maximum likelihood method of Ciliegi
et al. (\cite{ciliegi01}). The agreement 
is satisfactory with no systematic flux differences.
Deviations occur at low fluxes where either catalog is incomplete and for 
few outliers where algorithms disagree in splitting a peak into two sources
compared to one.

\begin{figure}
\centering
\includegraphics[width=\columnwidth]{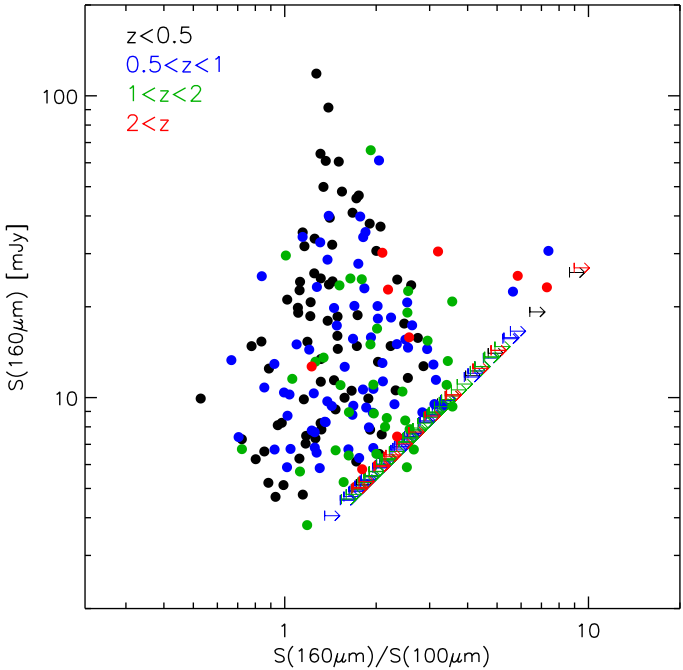}  
\caption{Color-magnitude diagram for 160~$\mu$m detected sources in the
GOODS-N field}
\label{fig:colormag}
\end{figure}

Figure~\ref{fig:colormag} shows a `color-magnitude' diagram for 160~$\mu$m
detected sources in GOODS-N. As expected, there is a tendency for higher
redshift sources to be redder in the 160/100~$\mu$m flux ratio, with 
considerable scatter due to measurement error and variation among the 
population. A detailed discussion of colors and SEDs is outside the
scope of this work, see also Elbaz et al. (\cite{elbaz10}), 
Magnelli et al. (\cite{magnelli10}), Hwang et al. (\cite{hwang10}) and
Dannerbauer et al. (\cite{dannerbauer10}) for first results on GOODS-N
spectral energy distributions.

\begin{figure}
\centering
\includegraphics[width=\columnwidth]{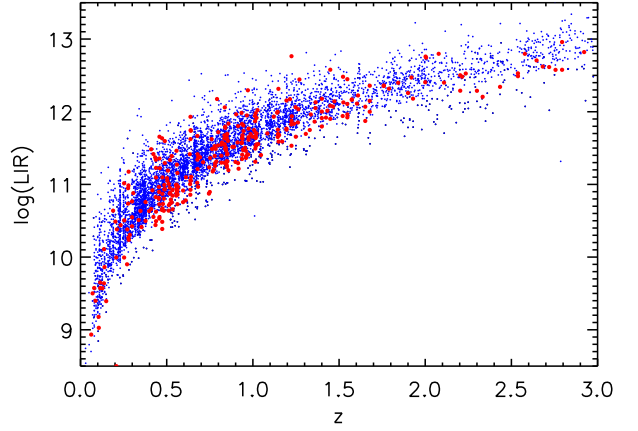}  
\caption{Red: Total infrared luminosity of GOODS-N sources detected
above 3$\sigma$ in the PEP SDP data. Blue: Sources from the GOODS-S and
COSMOS data illustrate the potential of the full PEP survey.}
\label{fig:lirz}
\end{figure}

Figure~\ref{fig:lirz} illustrates the potential of PEP to reach normal star
forming galaxies up to redshifts z$\sim$2. The GOODS-N science demonstration
phase data reach at z$\sim$2 infrared luminosities of 10$^{12}$\Lsun\  
(star formation rates about 100 solar masses per year). The GOODS-N SDP 
observation 
thus reach the star formation `main sequence' for massive galaxies at that
redshift (Daddi et al. \cite{daddi07a}). The GOODS-S data go 
deeper by a factor $\geq$2, and the COSMOS data provide the precious 
statistics at higher luminosities.

As an example of the PEP/PACS performance for deep surveys, the 
$>90\%$ completeness blind catalogs of the GOODSN and Abell 2218 SDP data 
will be released 
on the PEP website\footnote{http://www.mpe.mpg.de/ir/Research/PEP/index.php}.

\section{Overview of first science results of the PEP survey}
\label{sect:science} 

\subsection{The cosmic infrared background}
One of the most imminent tasks of a large far-infrared space telescope
is to resolve a large fraction of the cosmic infrared background
into its constituent galaxies. The CIB peaks at roughly 150~$\mu$m (e.g.
Dole et al. \cite{dole06}), making Herschel/PACS excellently suited to 
characterise the sources of its bulk energy output. Berta et al. 
(\cite{berta10}) used SDP observations of GOODS-N to derive 100 and 160~$\mu$m
number counts down to 3.0 and 5.7~mJy, respectively. Altieri et al.
(\cite{altieri10}) used strong lensing by the massive cluster Abell 2218
to push this limit down to 1 and 2~mJy, deeper by a factor $\sim$3 
for both wavelengths. Berta et al. 
(\cite{berta11}) use a complement of PEP blank field observations 
including the 
deep GOODS-S observations to reach a similar depth as Altieri et al. but 
with the larger statistics provided by the blank field observations. They 
also add 70~$\mu$m counts
overall reaching 1.1, 1.2, and 2.4mJy in the three PACS bands.
To this depth, 58\% (74\%) of the COBE CIB as quoted by 
Dole et al. 
(\cite{dole06}) is resolved into individually detected sources at 100 
(160~$\mu$m). These fractions reach 65\% (89\%) when including a P(D) analysis.

It is important to note that the direct COBE/DIRBE measurements have 
considerable 
uncertainties due to the difficulty of foreground subtraction. Similar to 
other wavelengths such as the $<$10keV X-rays (e.g. Brandt \& Hasinger 
\cite{brandt06}) we are reaching the point where the lower limits to
the cosmic background that are provided
by the integral of the resolved \herschel\ measurements are more 
constraining than the direct measurements. Quoting `resolved 
fractions' in reference to COBE hence starts to become problematic.

Because of the excellent multiwavelength coverage of the PEP fields,
spectroscopic or photometric redshifts can be assigned to the detected sources,
allowing us to determine the redshifts at which the CIB originates. To the
depth reached by Berta et al. (\cite{berta11}), half of the resolved CIB 
originates
at z$>$0.58, 0.67, 0.73 for wavelengths of 70, 100, and 160~$\mu$m, 
respectively. These redshifts are 
mild lower limits because they exclude faint sources that are not
individually detected by Herschel,
and naturally increase with wavelength, reflecting the dominant 
contributions by the far-infrared SED peak of increasingly distant objects.

Comparing the PEP counts and redshift distributions to backward 
evolutionary models tuned on pre-Herschel data, Berta et al. (\cite{berta11})
find for several models reasonable agreement with the total counts but 
more significant
mismatches to the observed redshift distributions. Clearly, models
have to be tuned in order to provide a satisfactory representation
of the most recent data including Herschel. 

The PEP GOODS-N SDP data have been used to derive first direct far-infrared
based luminosity functions up to z$\sim$2--3 
(Gruppioni et al. \cite{gruppioni10}). Strong evolution of the comoving 
infrared luminosity density (proportional to the star formation rate density)
is found, increasing with redshift as $(1+z)^{3.8\pm 0.3}$ up to z$\sim$1.
Global classification of the SEDs assigns to most objects either a 
starburst-like SED or an SED that is
suggesting a modest AGN contribution.

\subsection{`Calorimetric' far-infrared star
formation rates and star formation indicators}

Star formation rates are one of the key measurables in galaxy evolution 
studies. `Calorimetric' rest frame far-infrared measurements are 
the method of choice for massive and in particular dusty galaxies but have been
out of reach for typical high-z galaxies in the pre-Herschel situation. 
Typically, star formation rates were measured from the rest frame ultraviolet, 
mid-infrared, submm/radio, or a combination of those methods. All of them
require certain assumptions. Extracting a star formation rate from the
rest frame ultraviolet continuum involves breaking the degeneracies 
between the star
formation history and obscuration, and involves assumptions about the 
dust extinction law and/or geometry of the obscuring dust. Extrapolation
from the mid-infrared or submm/radio requires the adoption of SED templates
or assumptions that were derived locally but insufficiently tested at high
redshift.

Nordon et al. (\cite{nordon10}) focussed on massive z$\sim$2 normal 
star forming galaxies that
are currently the subject of intense study towards the role of secular
and merging processes in their evolution (e.g. F\"orster Schreiber et al.
\cite{foerster09}). They found ultraviolet-based star formation rates 
reasonably confirmed. At the rest frame optical bright end 
of these bright star forming galaxies, UV SFRs are
overpredicting the calorimetric measurement by a factor 2 only. In contrast,
a significant
overprediction by a factor $\sim$4--7.5 was found when extrapolating from
the 24~$\mu$m flux assuming that the local Universe 
Chary \& Elbaz (\cite{chary01})
templates apply {\em for the given mid-IR luminosity}. This is in line with
the encompassing SED studies of Elbaz et al. (\cite{elbaz10}) and Hwang
et al. (\cite{hwang10}) which, combining PEP and HerMES as well as local Akari
data,
found that this overprediction by 24~$\mu$m extrapolation sets in at z$\sim$1.5,
and that the FIR SED temperatures of high z galaxies are modestly colder 
than their local equivalents of similar total infrared luminosity.

There are two ways of looking at these results. Comparing at same total 
IR luminosity local galaxies with z$\sim$2
galaxies, the z$\sim$2 `overprediction' when 
extrapolating from the mid-IR could either be due to an enhanced ratio of the
PAH complex to the FIR peak, or due to a boosting of the mid-infrared
emission via the continuum of a possibly obscured AGN
(see also Papovich et al. \cite{papovich07} and Daddi et al. \cite{daddi07b}).
While most likely both factors are at work to some level, the setting 
in of the 24~$\mu$m
overprediction at z$\sim$1.5 as well as Spitzer spectroscopy of mid-IR
galaxies
at these redshifts (Murphy et al. \cite{murphy09}, Fadda et al.
\cite{fadda10}) argues for a dominant role of stronger PAH. This is
established
by Nordon et al. \cite{nordon11} who show that z$\sim$2 GOODS-S galaxies with
large 24~$\mu$m/FIR ratios indeed have PAH dominated spectra in the 
Fadda et al. (\cite{fadda10}) sample. 

It is important to put this into the perspective of other properties of 
$\lir\sim 10^{12}\Lsun$ local and z$\sim$2 galaxies. Locally, such an object
is a classical interacting or merging `ultraluminous infrared galaxy' (ULIRG,
Sanders \& Mirabel \cite{sanders96}) 
with star formation concentrated in a few hundred parsec sized region. 
Such star formation rates are well above the
local `main sequence' (Brinchmann et al. \cite{brinchmann04}). At z$\sim$2, 
objects with the same star formation
rates can be on the main sequence (Daddi et al. \cite{daddi07a}) and are 
often not interacting/merging but
massive, turbulent disks with star formation spread out over several kpc
(e.g. Genzel et al. \cite{genzel08}, Shapiro et al. \cite{shapiro08}, 
F\"orster Schreiber et al. \cite{foerster09}).
It is hence not surprising that SED templates calibrated on local compact
mergers fail to reproduce the more extended star formation at z$\sim$2.
The locally defined connotations of the `ULIRG' term, beyond its basic
definition  by IR luminosity, obviously do not apply automatically at 
high redshift.

Rodighiero et al. (\cite{rodighiero10}) use the direct Herschel 
far-infrared star 
formation rates in the PEP GOODS-N field to characterize the evolution of the 
Specific Star Formation Rate (SSFR)/mass relation up to z$\sim$2. They find a 
steepening from an almost flat local relation to a slope of -0.5 at z$\sim$2.

Concerning the most highly star forming high redshift objects, Magnelli et al.
(\cite{magnelli10}) combine the GOODS-N and Abell 2218 PACS data with submm 
surveys. They use PACS to sample the Wien side of the far-infrared SED of 
submm galaxies and optically faint radio galaxies with accurately known 
redshifts. The directly measured dust
temperatures and infrared luminosities are in good agreement with estimates
that are based on radio and submm data and are adopting the local universe 
radio/far-infrared correlation (see also the dedicated PEP/HerMES test of 
the high-z radio/far-infrared correlation by Ivison et al. (\cite{ivison10}). 
This confirmation of huge O(1000\Msun yr$^{-1}$) star formation rates for SMGs
is in support of a predominantly merger nature of SMGs, since such star 
formation rates are very hard to sustain with secular processes
(e.g. Dav\'e et al. \cite{dave10}).

\subsection{The role of environment} 

PEP data cover a wide range of environments. As first steps,  
Magliocchetti et al. (\cite{magliocchetti11}) derive correlation
functions and comoving correlation lengths at z$\sim$1 and 2 from the
GOODS-S data which also provide evidence for most infrared bright 
z$\sim$2 galaxies in this field residing in a filamentary structure.
At z$\sim$1, Popesso et al. (\cite{popesso11}) observe a reversal of the
local star formation rate - density relation which is linked to the 
presence of AGN hosts that are favoring high stellar masses, dense 
regions and high star formation rates.

\subsection{Studies of individual galaxy populations} 

Dannerbauer et al. (\cite{dannerbauer10}) use the $\lesssim 10\arcsec$ beam
PEP maps to verify the identifications of (sub)millimeter galaxies and test
the potential of PACS colors and mid-infrared to radio photometric redshifts
for these objects. Santini et al. (\cite{santini10}) use the Herschel 
points to break the degeneracy between dust temperature and dust mass that 
is inherent to submm-only data, and derive high dust masses for a sample 
of submillimeter galaxies. Comparison of these high dust masses with the 
relatively low gas phase metallicities either implies incompletely understood
dust properties or a layered structure which is combining low metallicity 
visible outer regions with a highly obscured interior.

Magdis et al. (\cite{magdis10}) use a stacked detection of z$\sim$3 Lyman
break galaxies to constrain their far-infrared properties. Similar to the 
SMG/OFRG study of Magnelli et al. (\cite{magnelli10}), this work highlights the
selection effects in the luminosity / dust temperature plane that are imposed
by groundbased submm detection. Bongiovanni et al. (\cite{bongiovanni10})
study Lyman $\alpha$ emitters in the GOODS-N field. PACS detections of 
part of these
galaxies are evidence for overlap with the dusty high redshift galaxy 
population.

\subsection{AGN-host coevolution: Two modes?}

Within the complex task of disentangling the coevolution of AGN and their 
host galaxies, and the role of AGN feedback, rest frame far-infrared 
observations 
provide a unique opportunity to determine host star formation rates. This
rests on the assumption that the host dominates over the AGN in the rest
frame far-infrared. This assumption is motivated by determinations of the 
intrinsic 
SED of the AGN proper, which is found to drop towards the far-infrared
(e.g. Netzer et al. \cite{netzer07}, Mullaney et al. \cite{mullaney11}). 
Compared to previous attempts
from the submm (e.g. Lutz et al. \cite{lutz10}) and Spitzer (e.g. 
Mullaney et al. \cite{mullaney10}), Herschel performance provides a 
big step forward. Shao et al. (\cite{shao10}) use the PEP GOODS-N 
observations and the Chandra Deep Field North to map out the host 
star formation of AGN of different redshift and AGN luminosity.
The host far-infrared luminosity of AGN with $\lhard\approx 10^{43}\ergs$
increases
with redshift by an order of magnitude from z=0 to z$\sim$1, similar to the
increase with redshift in star formation rate of inactive massive galaxies. 
In contrast,
there is little dependence of far-infrared luminosity on AGN luminosity, for
$\lhard\lesssim 10^{44}\ergs$ AGN at z$\gtrsim$1. 
In conjunction with properties of local and luminous high-z AGN, this suggests
an interplay between two paths of AGN/host coevolution. A correlation of AGN
luminosity and host star formation for luminous AGN 
reflects an evolutionary connection, likely via merging.
For lower AGN luminosities, star formation is similar to that in non-active
massive galaxies and shows little dependence on AGN luminosity.
The level of this secular, non-merger driven star formation increasingly
dominates over the correlation at increasing redshift.

\section{Conclusions}
Deep \herschel\ far-infrared surveys are a powerful new tool for the study of
galaxy evolution and for unraveling the constituents of the cosmic infrared 
background. We have described here the motivation, field selection, observing
strategy and data analysis for the PEP guaranteed time survey. Science
demonstration phase data of GOODS-N and Abell 2218 are discussed to illustrate
the performance of Herschel-PACS for this science. The wide range of initial
science results from the PEP data is briefly reviewed.

\begin{acknowledgements}
We thank the referee for helpful comments. PACS has been developed 
by a consortium of institutes led by MPE
(Germany) and including UVIE (Austria); KUL, CSL, IMEC (Belgium); CEA,
OAMP (France); MPIA (Germany); IFSI, OAP/OAT, OAA/CAISMI, LENS, SISSA
(Italy); IAC (Spain). This development has been supported by the funding
agencies BMVIT (Austria), ESA-PRODEX (Belgium), CEA/CNES (France),
DLR (Germany), ASI (Italy), and CICYT/MCYT (Spain).
\end{acknowledgements}

\end{document}